\documentclass[twocolumn, amsmath,superscriptaddress, amsfonts,prb]{revtex4}

\usepackage{multirow}
\usepackage{amsmath}
\usepackage{graphicx}
\usepackage{epstopdf}
\usepackage{dcolumn}
\usepackage{bm}

\begin{document}

\title{Evidence of rapid tin whisker growth under electron irradiation}
\author{A.C. Vasko}\affiliation{Department of Physics and Astronomy, University of Toledo, Toledo, OH 43606, USA}
\author{G. R. Warrell}\affiliation{Department of Radiation Oncology, University of Toledo Health Science Campus,
Toledo, Ohio 43614, USA}

\author{E. Parsai}\affiliation{Department of Radiation Oncology, University of Toledo Health Science Campus,
Toledo, Ohio 43614, USA}
\author{V. G. Karpov}\affiliation{Department of Physics and Astronomy, University of Toledo, Toledo, OH 43606, USA}
\author{Diana Shvydka}\affiliation{Department of Radiation Oncology, University of Toledo Health Science Campus,
Toledo, Ohio 43614, USA}\email{diana.shvydka@utoledo.edu}

\date{\today}

\begin{abstract}

We have investigated the influence of electric field on tin whisker growth. Sputtered tin samples were exposed to electron radiation, and were subsequently found to have grown whiskers, while sister control samples did not exhibit whisker growth. Statistics on the whisker properties are reported. The results are considered encouraging for substantiating an electrostatic theory of whisker growth, and the technique offers promise for examining early stages of whisker growth in general and establishing whisker-related accelerated life testing protocols.

\end{abstract}

\maketitle

Metal whiskers (MW) are hairlike protrusions that can grow on surfaces of many technologically important metals, for example, tin and zinc. MW across leads of electric equipment cause short circuits raising reliability concerns. The nature of MW remains a mystery after nearly 70 years of research. \cite{galyon2003,brusse2002,davy2014,zhang2004,tu2005,bunian2013} Procedures for MW mitigation are lacking; neither are there accelerated life testing protocols helping to predict their development. In spite of versatile information about MW, \cite{NASA} there is almost no attention to them from the physics community.

Recently, tin whisker reliability issues have been significantly aggravated, after the previously used mitigation recipe of introducing lead into tin was effectively eliminated by the environmental Restriction of Hazardous Substances (RoHS) directive adopted in the European Union since 2002. Use of lead-free tin-based solders has consequently spilled worldwide through production standards. This has endangered multiple industries that rely on tin soldering, including aviation, automotive, and medical device industries. Hence, the need for understanding, predicting, and mitigating MW becomes even more urgent.

Here, we describe the physical effect of rapid tin whisker growth under the electron beam of a  medical linear accelerator. The underlying motivations are that (1) this effect can be a potential accelerated life test for whisker-related reliability issues; (2) to our knowledge, the factor of ionizing radiation on whisker growth has never been considered, which overlooks whisker-related issues in a wide range of applications covering aerospace, particle accelerators, and medical equipment (such as pacemakers and defibrillators) under radiation; (3) this effect can shine new light on the physics behind whisker growth.

The possibility of electron beam effects on MW development [for example, Scanning Electron Microscope (SEM) beams] was predicted by a recently proposed electrostatic theory of metal whiskers \cite{karpov2014,karpov2015} attributing MW to random electric fields in the near-surface regions of imperfect metals. Moreover, a recent paper \cite{umalas2015} reported strongly accelerated silver MW growth under SEM beams. Based on Refs. \onlinecite{karpov2014,karpov2015,umalas2015}, our approach here is aimed at creating strong electric field perpendicular to tin film surface by charging the sample substrate under the electron beam of a medical linear accelerator.

Note that electric charging appears to be a major result of radiation for the present case of sub-micron thin tin films. Indeed, the probability of radiation defect creation in such films is negligibly small for high energy ($\sim 10$ MeV) electrons which have projected range on the order of centimeters. \cite{friedland2010,vineyard1956}

So far, observations of the electric bias effects on whisker growth have remained scarce and inconclusive.  Two groups have demonstrated rapid whisker growth through the use of high current density in tin films \cite{liu2004,crandall}, but did not isolate possible electric field effects. Other publications reported no bias effect on whisker growth \cite{hilty2005} and even a negative effect of bias-suppressed whiskering. \cite{ashworth2013} Our results below provide additional data on the possibility of electric field effects on whisker development.

Our approach here is consistent with multiple studies indicating the presence of strong electric fields due to material charging under electron beams. These studies mostly deal with the conditions of vacuum characteristic of SEM beams, \cite{melchinge1995,song1996,touzin2006,mizuhara2002} but some include ambient air effects. \cite{wilson2013,lai1989} A significant factor provided by the latter is air ionization, which establishes an efficient current flow and determines the electric potential and field distributions around the sample and their time decay after beam removal.  This is critical for the present study as well.

\begin{figure}[t!]
\includegraphics[width=0.44\textwidth]{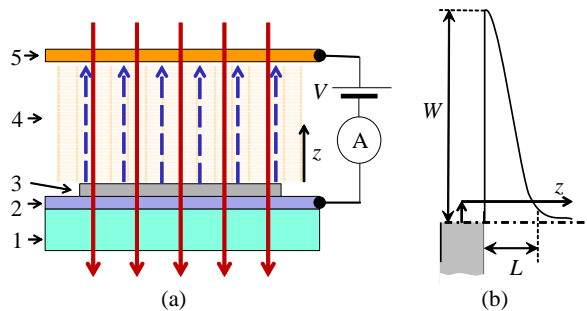}
\caption{(a) Sketch of the experimental setup: downward arrows represent the primary electron beam; upward arrows show the current of electrons from the tin film. The layers represent: 1 - glass substrate, 2 - conductive oxide layer, 3 - tin film, 4 - insulating spacer between two electrodes filled with plastic transparency (slide) films, 5 - second (copper or aluminum foil) electrode. A and V represent the ammeter and power source, $z$ is the axis perpendicular to the tin electrode. (b) Band diagram with tin layer showing the Fermi level (dash-dotted line), the states filled with electrons (gray), and the work function $W$. The electron tunneling current is shown by the $z$-axis arrow. \label{Fig:setup}}
\end{figure}

Our tin films were sputter coated on Pilkington TEC15 substrates (soda lime glass with nominal 15 $\Omega / \square$ sheet resistance SnO$_2$:F coating; see Fig. \ref{Fig:setup}).  The SnO$_2$:F coating allows for easy electrical connection at all processing stages, and its surface roughness allows for greater film adhesion.  The insulating glass substrate serves to accumulate a net charge from the incident radiation.
Sputtering was done at room temperature from a pure 0.25 inch tin target, in a 20 mTorr working pressure atmosphere of argon gas, using 50 watts RF power. The deposited film thickness of 0.15 $\mu$m is below the micron range of efficient whisker growing film thickness. \cite{galyon2003,brusse2002,davy2014,zhang2004,tu2005,bunian2013} In our experiments, a sample of tin film on the glass substrate was placed on a 2 mm thick acrylic pedestal left ungrounded to facilitate electric charge accumulation under the incident electron beam.

The 6 MeV electron beam of a Varian TrueBeam (Varian Medical Systems, Palo Alto, California) medical linear accelerator, operated in service mode (the machine diagnostic and research mode), was used to irradiate the sample. The  electron fluence rate and their transmitted energy are estimated to be respectively  $G\approx 10^9$ electrons/(cm$^2$-s) and $w\approx 2$ MeV/electron. The irradiation was conducted in two 10-hour rounds, each spread out over five daily sessions two hours long each. The sample structure is represented by layers 1,2,3 in Fig. \ref{Fig:setup} showing also the direction of irradiation.

\begin{figure}[t!]
\includegraphics[width=0.44\textwidth]{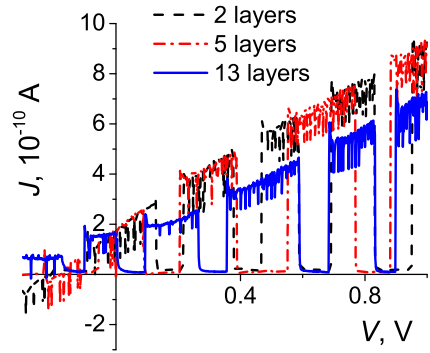}
\caption{Current-voltage characteristics of the structure depicted in Fig. \ref{Fig:setup} for three different insulating spacers between the tin and second electrode. Note that the machine was repeatedly turned on and off, leading to the gaps in the plots. \label{Fig:current}}
\end{figure}

To evaluate the field strength, a second electrode was added, as illustrated in Fig. \ref{Fig:setup} (a); it was sufficiently thin ($\lesssim 0.1$ mm) that the beam characteristics were practically unchanged. To avoid shunting, the inter-electrode gap was filled (and its thickness defined) with several layers of plastic sheets of thickness $< 0.1$ mm each. The current voltage characteristics corresponding to  different numbers of such sheets were measured with a Keithley  dual channel source meter (Keithley Instruments, Inc., Cleveland, Ohio) as shown in Fig. \ref{Fig:current}. During each measurement, the beam was turned on and off several times. Note that neither the second electrode nor plastic sheets were present during the exposure leading to whisker growth.

\begin{figure}[b!]
\includegraphics[width=0.44\textwidth]{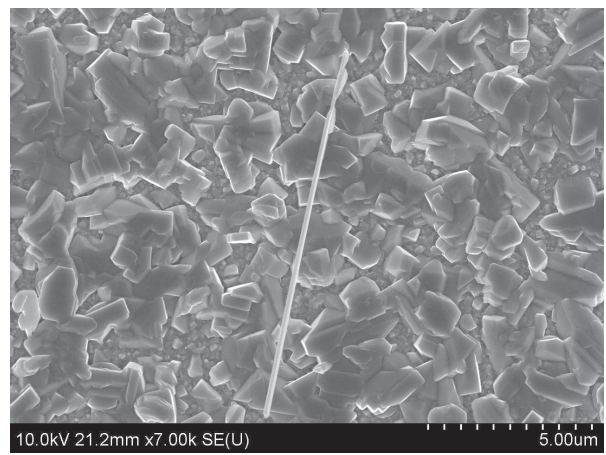}
\caption{An example photograph of a single whisker grown after 10 hours of electron beam exposure.\label{fig:whisker}}
\end{figure}

\begin{figure}
\includegraphics[width=0.45\textwidth]{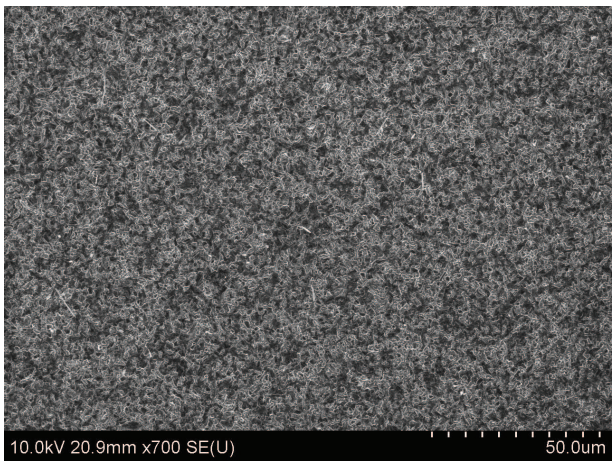}
\caption{A representative wide overview showing many whiskers (better seen under higher magnification).  10 whiskers were counted in this region corresponding to the density of 500 whiskers$/$mm$^2$ \label{fig:overview}}
\end{figure}

The following observations are important for evaluating the electric filed distribution. (i) The measured current sense corresponds to the electron flow depicted in Fig. \ref{Fig:setup} (a), i. e. negative charge is acquired by the structure. (ii) The short circuit ($V=0$) current does not depend on the gap thickness and the number of layers inserted between the electrodes. (iii) The slope of current voltage characteristics does not depend on the number of layers until it is very large and then shows only weak dependence (the case of 13 layers in Fig. \ref{Fig:current}). (iv) Changing the electron fluence rate by an order of magnitude results in  an insignificant (factor of 2) current change. (v) No delays between voltage and currents were observed, and the system behavior was not altered by intermittent grounding, i. e. the structure did not retain any significant charge upon beam removal.

To interpret these observations, we note that the sign of the charge acquired by a dielectric (particularly, glass) sample under electron irradiation depends on the sample composition, electron stopping range and grounding. \cite{miyake2007,wilson2013} The 6 MeV electron stopping range of about 2 cm in a glass is considerably greater than the sample and pedestal integral thickness of $\lesssim 1$ cm; hence the majority of primary (originating from the beam itself) electrons pass completely through the experimental setup. This creates large numbers of electron-hole pairs as well as radiation-induced conductivity in the glass. \cite{wilson2013} Some of these electrons and holes can escape through the sample surface leaving unbalanced positive or negative charges depending on the material composition \cite{miyake2007} and interfacial conditions; our data show that the sample is charged negatively.

In our setup without dedicated grounding, electric charge exchange is achieved through the air ionization \cite{lai1989} providing sufficient ionic conduction between the sample and the accelerator. Furthermore, the presence of significant ion concentration screens the electric field of the charged sample forming a barrier shown in Fig. \ref{Fig:setup} (b); its height is equal the tin work function $W=4.7$ eV. The barrier shape is determined by the distribution of  ions in the air (the plastic layers do not have enough mobile charges to contribute). The barrier thickness $L$ is considerably smaller than the inter-electrode gap: using the standard value of the air dielectric strength, $E\approx 30,000$ V/cm, one can estimate $L=W/E\approx 2$ $\mu$m. The barrier-dominant resistance then determines the value of short circuit current accommodating practically the entire external voltage applied and making the current voltage slope insensitive to the gap structure, which explains the above observations (ii) - (iii). The observation (iv) confirms that most of the created electron hole pairs recombine in the structure. Finally, (v)  can be explained by the radiation-induced conductivity \cite{dennison2009} making the corresponding RC times shorter than the characteristic times of our experiments.

After first 10 hours of electron beam exposure, the sample was examined, for whisker formation by a Hitachi S-4800 SEM.  Measured whisker length was taken as the entire whisker, including kinks if necessary (as opposed to merely the straight-line distance between the beginning and end).

We found that the irradiation procedure was successful in growing whiskers.  Moreover, control samples not exposed to the radiation  grew no detectable whiskers whatsoever over the same time period.

Fig. \ref{fig:whisker} shows an example of single whisker, illustrating the high aspect ratio of this feature in comparison to the prevailing grain structure.
Fig. \ref{fig:overview} shows a larger region of the sample, illustrating the density of whiskers on the sample.

\begin{figure}[tbh]
\includegraphics[width=.40\textwidth]{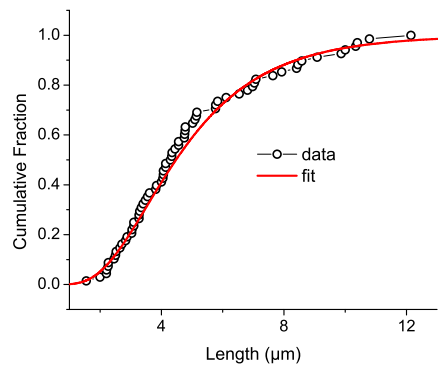}
\caption{The log-normal fit of the whisker length statistical distribution. \label{fig:statfit}}
\end{figure}

Previous work by other groups has suggested that tin whisker length is described by a log-normal distribution \cite{fang2006,panashchenko2009,susan2013}.  Figure \ref{fig:statfit} shows a satisfactory fit of measured whisker lengths  with the cumulative probability of a log-normal distribution.    The fitting parameters correspond well to the directly determined values of average length of 5.0$\mu$m with standard deviation of 2.5$\mu$m. The average whisker density (whiskers per area) was determined to be (335$\pm$32) mm$^{-2}$.

The sample was given then another 10 hours of radiation before the second measurement of whisker length. Assuming that whisker growth was due entirely to the 20 hours of irradiation time, the whisker growth rate was 0.7$\pm$0.3 \AA /s, which is in line with whisker growth induced by mechanical stress in some studies, and is at least an order of magnitude greater than what is reported for spontaneous whisker growth ($\sim 0.0002-0.1$ \AA /s) \cite{fang2006,galyon2003}. It should be emphasized again that the tin film thickness used in this study is generally considered not to be prone to whisker formation; hence, the field effect can provide even more significant acceleration of whisker growth rate.

We consider these results to be promising first steps towards clarifying the role of electric field and trapped charges in tin whisker growth.  More work is called upon including solder representative (micron thickness electroplated) tin samples, variations in radiation dose and fluence, etc. to investigate whisker growth as a function of time, field strength, film thickness, and alloy composition/morphology  under electric field; other groups efforts could strongly facilitate this ambitious effort.

As a practically important result, this work shows that radiation can have a significant effect on tin whisker growth that needs to be taken into account in medical device, space, and some other industries. Because ionizing radiation is a well-controllable tool, it can be developed to become an accelerated life test in whisker-related reliability problems.

{\it Acknowledgement}. Microscopy for this study was done with equipment at the Center for Materials and Sensor Characterization of the University of Toledo.


\begin{thebibliography}{99}

\bibitem{galyon2003}G. T. Galyon, Annotated Tin Whisker Bibliography And Anthology, IEEE Transactions on electronics Packaging Manufacturing, {\bf 28}, 94 (2005);\\ \url{http://thor.inemi.org/webdownload/newsroom/TW_}\\\url{biblio-July03.pdf}

\bibitem{brusse2002}J. Brusse, G. Ewell, and J. Siplon, Tin Whiskers: Attributes and Mitigation, Capacitor and Resistor
Technology Symposium (CARTS), March 25-29,  pp. 68-80, (200).
\bibitem{davy2014}G. Davy, private communication, October 2014; quated in Ref. \onlinecite{karpov2015}.

\bibitem{zhang2004}Y. Zhang, Tin Whisker Discovery and Research, in {\it Soldering in Electronics}, Edited by K. Suganuma, Marcel Dekker, Inc. p. 121 (2004)
\bibitem{tu2005}K.N. Tu, J.O. Suh, and Albert T. Wu, Tin Whisker Growth on Lead-Free Solder Finishes, in {\it Lead-Free Solder Interconnect Reliability}, Edited by D. Shangguan, ASM International, p. 851 (2005).
\bibitem{bunian2013}D. Bunyan, M. A. Ashworth, G. D. Wilcox, R. L. Higginson, R. J. Heath, C. Liu, Tin whisker growth from electroplated finishes – a review, Transactions of the institute of metal finishing, {\bf 91}, 249-259 (2013).
\bibitem{NASA}NASA Goddard Space Flight Center Tin Whisker Homepage, website \url{http://nepp.nasa.gov/whisker}.\\Bibliography for Tin Whiskers, Zinc Whiskers, Cadmium Whiskers, Indium Whiskers, and Other Conductive Metal and Semiconductor Whiskers,
John R. Barnes; \url{http://www.dbicorporation.com/whiskbib.htm}
    \bibitem{karpov2014}V.G.Karpov,	Electrostatic Theory of Metal Whiskers, Phys. Rev. Applied, {\bf 1}, 044001 (2014).
\bibitem{karpov2015}V.G.Karpov, Electrostatic Mechanism of Nucleation
and Growth of Metal Whiskers, SMT Magazine, February 2015, p. 28.\\ \url{http://iconnect007.uberflip.com/i/455818/44}

\bibitem{umalas2015} M. Umalas, S. Vlassov, B. Polyakov, L. M. Dorogin, R. Saar,
I. Kink, R. Lohmus, A. Lohmus, A. E.Romanov, Electron beam induced growth of silver nanowhiskers, J. Cryst. Growth, {\bf 410}, 63 (2015).

\bibitem{friedland2010}E. Friedland, Radiation Damage in Metals, Critical Reviews in Solid State and Materials Sciences, {\bf 26}, 87 (2010).
\bibitem{vineyard1956}G. H. Vineyard, Theory and Mechanism of Radiation Effects in Metals, Nuclear Metallurgy, IMD Special Report Series No. 3, AIME, 1 (1956).

\bibitem{liu2004}S. H. Liu, C. Chen, P. C. Liu,  and T. Chou,  Tin whisker growth driven by electrical currents, J. Appl. Phys. {\bf 95}, 7742 (2004).


\bibitem{crandall}E. R. Crandall, Factors governing thin whisker growth, Ph. D. Thesis, Auburn, Alabama (2012); \url{http://ldfcoatings.com/articles/ErikaCrandall.pdf}.
\bibitem{hilty2005}R. D. Hilty, N. Corman, and H. Herrmann, Electrostatic fields and current-flow impact on whisker growth, IEEE Trans. Elec. Packaging Manufacturing, {\bf 28}, 75-81 (2005).
\bibitem{ashworth2013} M. A. Ashworth, G. D. Wilcox, R. L. Higginson, R. J. Heath, and C. Liu, Effect of direct current and pulse plating parameters on tin whisker growth from tin electrodeposits on copper and brass substrates, Transactions of the  Institute of Metal Finishing, {\bf 91}, 260-268 (2013).
\bibitem{melchinge1995}A. Melchinge and S. Hofmann,
Dynamic double layer odel: Description of time dependent charging
phenomena in insulators under electron beam irradiation, J. Appl. Phys. {\bf 78}, 6224, (1995).

\bibitem{song1996}Z. G. Song, C. K. Ong, and H. Gong, A time-resolved current method for the investigation of charging ability
of insulators under electron beam irradiation, J. Appl. Phys. {\bf 79}, (1996).

\bibitem{touzin2006}M. Touzin, D. Goeuriot, C. Guerret-Piecourt, D. Juve, D.
Treheux, Electron beam charging of insulators: A self-consistent
ight-drift model, J. Appl. Phys., {\bf 99}, 114110 (2006).

\bibitem{mizuhara2002}Y. Mizuhara, J. Kato, T. Nagatomi, and Y. Takai, Quantitative measurement of surface potential and amount of charging
on insulator surface under electron beam irradiation, J. Appl. Phys. {\bf 92}, 6128 (2002).

\bibitem{wilson2013}G. Wilson, J. R Dennison, A. Evans, and J. Dekany, Electron Energy Dependent Charging Effects of Multilayered
Dielectric Materials, IEEE trans. on Plasama Science, {\bf 41}, 3536 (2103).

\bibitem{lai1989}S. T. Lai, An overview of electron and ion beam effects in charging and discharging spacecraft, IEEE Transactions on nuclear science, {\bf 36}, 2027, (1989).
\bibitem{miyake2007}H. Miyake, Y. Tanaka, T. Takada, Characteristic of charge accumulation in glass materials under electron beam irradiation, IEEE Trans. On Dielectrics And Electrical Insulation {\bf 14}, 520 (2007).
\bibitem{dennison2009}J. R. Dennison, J. Gillespie, J. Hodges, R. C. Hoffmann, J. Abbott, S. Hart, and A. W. Hunt, Temperature Dependence of Radiation Induced Conductivity in Insulators, Edited by: F. D. McDaniel, B. L. Doyle, {\it Application of Accelerators in Research and Industry},  Book Series: AIP Conference Proceedings   {\bf 1099}, 203  (2009)

\bibitem{fang2006}T. Fang, M. Osterman, M. Pecht, Statistical Analysis of Tin Whisker Growth, Microelectronics Reliability {\bf 46}, 846 (2006).
\bibitem{panashchenko2009}L. Panashchenko, Evaluation of environmental Tests for tin whisker assessment, MS Thesis, University of Maryland (2009). http://hdl.handle.net/1903/10021

\bibitem{susan2013}D. Susan, J. Michael, R. P. Grant, B. McKenzie \& W. G. Yelton, Morphology and Growth Kinetics of Straight and Kinked
Tin Whiskers, Metall and Mat Trans A, {\bf 44}, 1485 (2013).



\end{thebibliography}
\end{document}